\documentclass[12pt,preprint]{aastex}
\usepackage{times}

\shorttitle{Correlation between Kinetic and Magnetic Helicity}
\shortauthors{Gao, Zhao, \& Zhang}

\begin{document}
\title{Analysis on Correlations between Subsurface Kinetic Helicity
and Photospheric Current Helicity in Active Regions}

\author{Yu Gao\altaffilmark{1}, Junwei Zhao\altaffilmark{2}, and
         Hongqi Zhang\altaffilmark{1}}
\altaffiltext{1}{Key Laboratory of Solar Activity, National
Astronomical Observatories of China, Chinese Academy of Sciences,
Beijing, China } \altaffiltext{2}{W.~W.~Hansen Experimental Physics
Laboratory, Stanford University, Stanford, CA 94305-4085, USA}

\begin{abstract}

An investigation on correlations between photospheric current
helicity and subsurface kinetic helicity is carried out by analyzing
vector magnetograms and subsurface velocities for two rapidly
developing active regions. The vector magnetograms are from the {\it
SDO}/HMI ({\it Solar Dynamics Observatory} / Helioseismic and
Magnetic Imager) observed Stokes parameters, and the subsurface
velocity is from time-distance data-analysis pipeline using HMI
Dopplergrams. Over a span of several days, the evolution of the
weighted current helicity shows a tendency similar to that of the
weighted subsurface kinetic helicity, attaining a correlation
coefficient above 0.60 for both active regions. Additionally, there
seems to be a phase lag between the evolutions of the unweighted
current and subsurface kinetic helicities for one of the active
regions. The good correlation between these two helicities indicate
that there is some intrinsic connection between the interior
dynamics and photospheric magnetic twistedness inside active
regions, which may help to interpret the well-known hemispheric
preponderance of current-helicity distribution.

\end{abstract}


\section{Introduction}

\citet{Parker55} initiated a model for the cycle of magnetic field
in the sun with a mechanism named ``cyclonic motion", and
\citet{steen66} established the feasibility of AC or DC dynamos in
astrophysical environment. At that time, they formulated a
consistent explanation of the role of the cyclonic motions, which
was later called $\alpha$-effect \citep[e.g.,][]{Roberts71}.
\citet{Pouquet76} suggested that turbulence and magnetic fields were
common features of many celestial bodies and the $\alpha$-effect
contained both the kinetic and magnetic helicity. The correlation
between these two kinds of helicities has been under debate
\citep[e.g.,][]{Brand90, Brand95}. \citet{Kein83} gave an example of
$\alpha$-effect including zero kinetic helicity and the
$\alpha$-effect relied on the ensemble average of fluctuating
velocity and magnetic field. That is to say, the generation of
magnetic field requires current helicity but not necessarily the
kinetic helicity. The $\alpha$-effect is central to dynamo theory
and magnetic field generation, however, it lacked observational and
experimental supports over a long term \citep{Sokoloff07}. Not until
the recent 20 years, mirror asymmetries of magnetic field (magnetic
helicity) \citep{Seehafer, Pevtsov95, BZ98, HS04, Zhang10} and the
velocity (kinetic helicity) \citep{ZhKos03} were found in
observations. Apparently, frozen flow with the magnetic field in the
fluid of high magnetic Reynold number is a necessary condition for
the $\alpha$-effect in the framework of Parker's dynamo model. It
can be easily inferred that the variation of kinetic and current
helicity in the fluid might be a good manifestation of the freezing
process. Although the speculation of the depth of $\alpha$-effect is
still under debate, it is expected that some relevant characteristic
might be captured even near the photosphere.

Local helioseismology provides a unique tool to determine the
sub-photospheric flows of active regions. A statistical study by
\citet{Zhao04} showed that the subsurface kinetic helicity inside
active regions observed by the {\it Solar Heliospheric Observatory}
/ Michelson Doppler Imager ({\it SOHO}/MDI) observations seemed to
have a hemispheric preponderance,
 like what magnetic (or current) helicity observations had shown \citep{Pevtsov95, BZ98}. \citet{GZZ09}
analyzed the connection between the photospheric current helicity,
calculated from vector magnetograms observed by Huairou Solar
Observing Station, and the subsurface kinetic helicity measured from
MDI observations in 38 solar active regions.  Although there was an
opposite hemispheric trend between the sign of current helicity and
that of subsurface kinetic helicity near the solar surface, the
result did not support that the subsurface kinetic helicity had a
cause and effect relation with the photospheric current helicity at
the depth of 0-12 Mm. Similar result was reported by
\citet{Mauetal11} as well.

Now, {\it Solar Dynamics Observatory} / Helioseismic and Magnetic
Imager \citep[{\it SDO}/HMI;][]{Schetal12, Schou12} observations
provide an unprecedented opportunity to investigate the connection
between subsurface kinetic helicity and current helicity, as both
subsurface flow velocity and photospheric vector magnetic field are
available at the same time.
 Subsurface kinetic helicity can be computed from subsurface flow velocities, which are routinely processed through the HMI time-distance analysis pipeline \citep{Zhao12}. Photospheric current helicity is able to be computed from the photospheric vector magnetograms \citep{Hoeksema12}.
In this Letter, we focus our study on active regions NOAA AR~11158
and AR~11283, both of which have continuous observational coverage
that allows us to investigate the relationship between subsurface
kinetic helicity and current helicity over a longer period rather
than just a snapshot comparison as done in our previous study
\citep{GZZ09}. In \S2, we describe the procedure of observation and
data preparation for this analysis, and present our results in \S3.
We then summarize and discuss our results in \S4.

\section{Observation and Data Reduction}

HMI observes the full-disk Sun continuously, providing Doppler
velocity and line-of-sight magnetic field, among others, with a
45-sec cadence, and also vector magnetic field with a cadence of 12
min \citep{Schou12}. Each full-disk image has 4096 $\times$ 4096
pixels with a spatial resolution of 0.504 arcsec pixel$^{-1}$ (i.e.,
approximately, 0.03 heliographic degree pixel$^{-1}$ at the solar
disk center). Subsurface flow velocities are computed from HMI
Doppler-shift observations using time-distance helioseismology
data-analysis pipeline \citep{Zhao12}. In practice, to execute this
data-analysis pipeline, users provide the Carrington coordinate for
the interested area and time of the interested period, then the
pipeline code selects an area of roughly $30^{\circ} \times
30^{\circ}$ centered at the given coordinate, and a duration of 8
hours with the given time as the middle point. Normally, the
pipeline generates a subsurface flow field consisting 256 $\times$
256 pixels with a horizontal spatial sampling of 0.12$^{\circ}$
pixel$^{-1}$, and a number of depths covering from the photosphere
to 20 Mm in depth. 
We performed analysis in all depths, and found that the 
shallowest depth, i.e., 0 -- 1~Mm, gave the most consistent and robust 
results. Thus, in this Letter, we only present results obtained from this 
depth and leave analyses of deeper layers for future studies.
And also, a recent comparison of subsurface flow field at the depth
of 0 -- 1~Mm, obtained from this
time-distance data-analysis pipeline, and photospheric flows
obtained from the DAVE4VM technique \citep{sch08} found a reasonable
agreement in horizontal flow fields inside active regions
\citep{liu2012}. Vector magnetograms are also from HMI observations,
and these data have also been widely used in carrying out
quantitative analysis \citep[e.g.,][]{Sun12}.

\begin{figure}
\epsscale{.60} \plotone{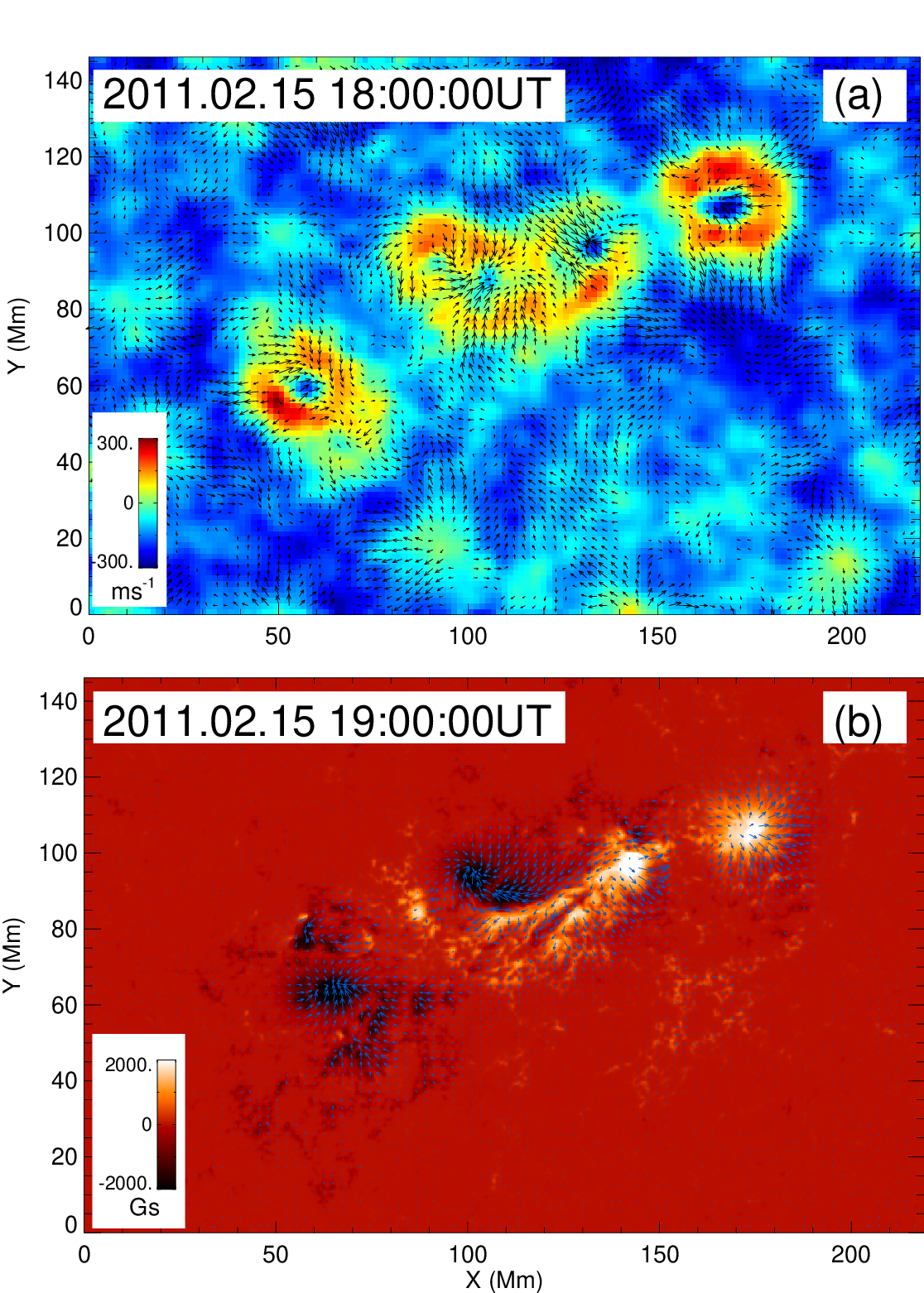} \caption{(a) Example of subsurface
velocity field of NOAA AR~11158 at a depth of 0 -- 1 Mm. Background
shows the vertical component of the velocity and arrows show the
horizontal components. The maximum vertical velocity is 285 m
s$^{-1}$ and the longest arrow represents a horizontal speed of
441.8 m s$^{-1}$. The spatial sampling of the vertical component is
2.016$^{\prime\prime}$$ \times $2.016$^{\prime\prime}$, and the
field of view is 3.02$^{\prime}$$ \times $2.02$^{\prime}$. (b)
Example of vector magnetogram of NOAA AR~11158, with white showing
positive polarity and black showing negative. The longest arrow
represents a transverse field of 2356~G, and the transverse field
lower than 300~G is not shown. } \label{fig1}
\end{figure}

\begin{figure}
\epsscale{.70} \plotone{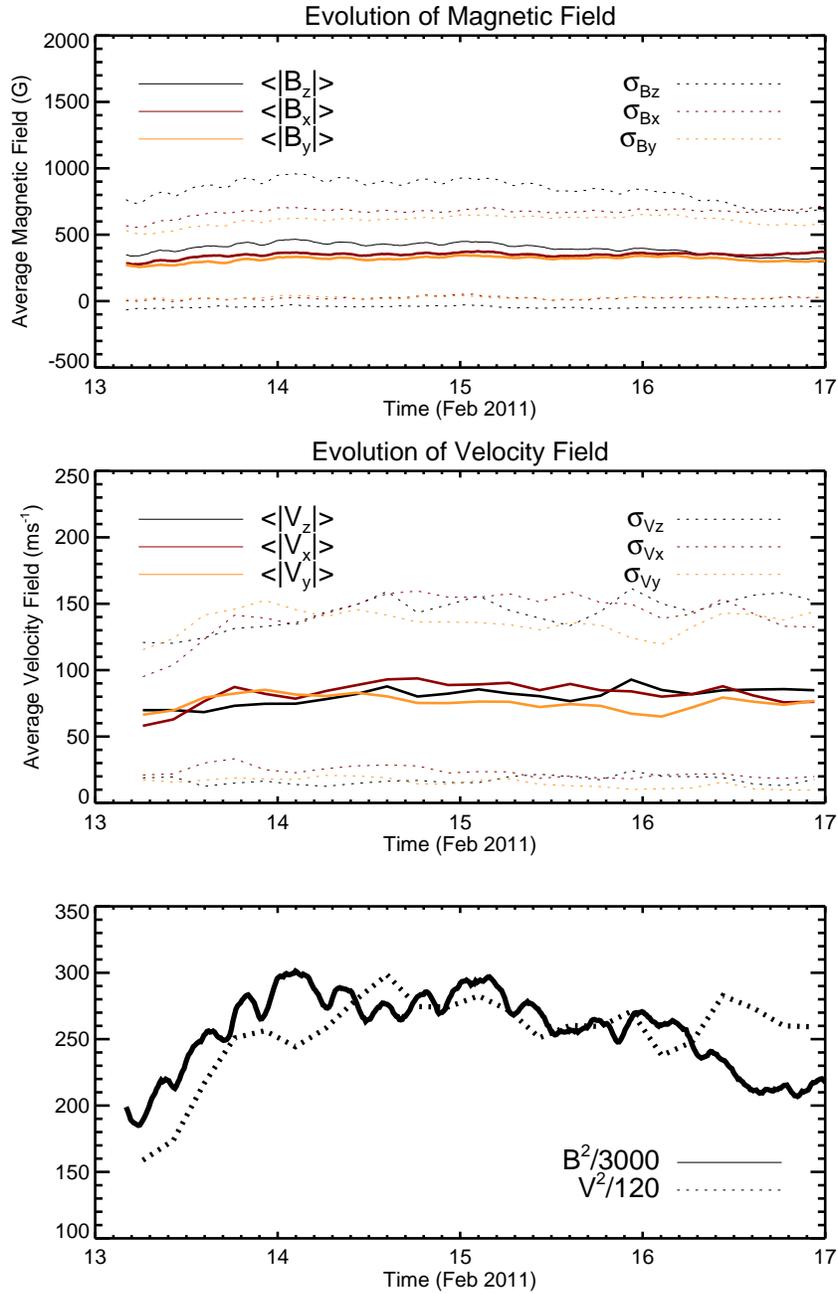} \caption{Temporal evolution
of the three components of the magnetic field ({\it top}), the
three components of the subsurface velocity field ({\it middle}) and 
the square of magnetic field and velocity ({\it bottom}).}
\label{fig2}
\end{figure}

NOAA AR~11158 was a rapidly developing active region, in which an
X~2.2 solar flare started at 01:44 UT and peaked at 01:56 UT of 2011
February 15. This active region, as well as the X-class solar flare
that occurred in it, were widely studied for different purposes
\citep[e.g.,][]{Sun12, Jetal12, liu2012}. Figure~\ref{fig1}a shows
an example of velocity field of AR 11158 at the depth of 0 -- 1~Mm,
with the background showing the vertical component of the flow and
arrows showing the horizontal components. Figure~\ref{fig1}b shows
an example of vector magnetogram of NOAA 11158 taken at a time
similar to the subsurface flow field. Figure~\ref{fig2} shows the
evolution of the averaged absolute values of three components of the
magnetic field and three components of the subsurface velocity,
respectively. It can be found that the magnitudes of the components
of the velocity and magnetic field remain relatively stable, and
there is no coherent or sudden change for any of these quantities.
Meanwhile, the fluctuations of these quantities are also small, as
shown in the curves of standard deviations.

Another active region, NOAA~AR~11283, is also studied. This region
was very flare productive, generating two C-class solar flares from
22:00~UT to 23:33~UT on 2011 September 5, an X2.1 flare that started
at 22:12 UT and peaked at 22:20 UT on September 6, an X1.8 flare
that started at 22:32 UT and peaked at 22:38 UT, and an M6.7 flare
on September 8 when the region was close to the west limb. As shown
in \S3, the same analysis on evolutions of photospheric current
helicity and subsurface kinetic helicity is applied on both AR~11158
and AR~11283, and similar results are achieved for both active
regions.

\section{Results}

The main purpose of this work is to study whether there is a
correlation between the photospheric current helicity and subsurface
kinetic helicity inside active regions. Current helicity is defined
as: $ H_c=\mathbf{B} \cdot (\nabla \times \mathbf{B})$. Similar to
what was employed by \citet{BZ98}, the averaged value of the
vertical component density of the current helicity, denoted as
$\langle H_c^z\rangle$, was used in this study. In addition, we also
compute the weighted $\langle H_c^z\rangle$, defined as $\langle
H_c^z\rangle$ divided by the total magnetic field strength, i.e.,
$\langle H_c^z\rangle / |\mathbf{B}^2|$. 
This parameter has the same dimension as the mean twist $\alpha$,
a commonly used proxy of magnetic helicity. However, $\alpha$ is defined 
as the curl of transverse magnetic field dividing the corresponding 
longitudinal field, hence carries a same sign as the weighted 
$\langle H_c^z\rangle$.
Similarly, kinetic helicity
is defined as $H_k = \mathbf{v} \cdot (\nabla \times \mathbf{v})$,
and we take the averaged value of its vertical component, $\langle
H_k^z\rangle$, and the value weighted by the speed, $\langle
H_k^z\rangle / |\mathbf{v}^2|$, for further studies. To compute both
values of $\langle H_c^z\rangle$ and $\langle H_k^z\rangle$, we only
use the areas where $|B_z| > 50$ Gs, and this is a popular approach
taken in many previous studies, e.g., \citet{BZ98}. 
The magnetograms are rebinned to the same resolution of the velocity 
map, and all quantities are averaged over the same spatial areas. 
In this Letter, we study the evolutionary relationship between $\langle
H_c^z\rangle$ and $\langle H_k^z\rangle$ inside the two
flare-productive active regions introduced above.

The top panel of Figure~\ref{fig3} shows evolutions of the weighted
current-helicity density and weighted kinetic-helicity density
obtained at a depth of 0 -- 1 Mm for AR~11158. The time series from
2011 February 13 to 17 are analyzed. The weighted current-helicity
density with a 12-min cadence is averaged over a 4-hr period in
order to compare with the weighted kinetic-helicity density, which
is computed from the subsurface velocity obtained from an 8-hr data
sequence with a 4-hr time step. Error bars are plotted for the
current helicity with corresponding standard deviations. The two
helicity curves show very similar varying tendencies and the
correlation coefficient attains 0.67. Furthermore, we separate the
data series into two sections. For the decreasing phase before
06:00UT of 2011 February 14, the correlation coefficient between the
kinetic-helicity density and the current-helicity density is as high
as 0.84. But for the rising phase after 06:00UT of 2011 February 14,
the correlation coefficient drops to 0.52. The decrease of the
correlation may be related to the X2.2 flare event that occurred at
01:44 UT of 2011 February 15. 
For the calculation of these correlation coefficients, the 
included data points n (correlation coefficients r) are 22 (0.67),
7 (0.84), and 15 (0.52), thus the degrees of freedom 
are 20, 5, and 13, respectively. The corresponding critical values 
of the Pearson correlation coefficients for a significance level of 
0.95 are 0.423, 0.754, and 0.514, respectively. Therefore, the correlation
coefficients from our calculations are all highly significant.

The bottom panel of Figure~\ref{fig3} shows the evolutions of the
unweighted current-helicity density and unweighted subsurface
kinetic-helicity density. Different from the similar evolutionary
tendency of the weighted parameters, the two unweighted parameters
seem to evolve out of phase. As shown in this panel, the unweighted
kinetic helicity seems to have a decreasing phase about 8~hr earlier
than that of the current helicity before 10:00 UT of February 14,
and have an increasing phase about 4~hr behind that of the current
helicity after 18:00 UT of February 14. 
The bottom panel of Figure~\ref{fig2} shows the evolution of 
$B^2$ and $v^2$. These two quantities seem to evolve out of phase
at the beginning and more in phase after mid-day of February 14. This 
may account for the evolutionary difference in the weighted and 
unweighted kinetic helicty.

The snapshots in Figure~\ref{fig4} show maps of the weighted and unweighted
$\langle H_k^z\rangle$ as well as the weighted and unweighted
$\langle H_c^z\rangle$ for some selected periods. It can be found
that both of the kinetic-helicity parameters are more fragmented
than the current-helicity parameters, but there is not clear
correspondence between the sign distributions of the kinetic
helicity and the current helicity. Besides, we also plot 
weighted $\langle H_c^z\rangle$ and unweighted $\langle H_c^z\rangle$ after binning
down to match the spatial resolution of $\langle H_k^z\rangle$. Furthermore, we compute the correlation between the maps of weighted (unweighted) current and 
kinetic helicity with the same resolution. The 
values are 0.019 and 0.028 respectively.

For the other active region NOAA~AR~11283, the evolutions of both
current helicity and subsurface kinetic helicity, both weighted and
unweighted, are shown in Figure~\ref{fig5}. Similar to the results
for AR11158, the evolutionary curves of the weighted parameters have
a better correlation, which is 0.62 for this case, than those of the
unweighted parameters. However, for the unweighted parameters, we do
not see the phenomenon of that the subsurface kinetic helicity
evolves ahead of the current helicity before the flare and behind
the current helicity after the flare, like what is observed in
AR~11158. 
The evolution of $B^2$ and $v^2$ for this active region does not 
show any phase lag either.

\section{Discussion}

We have studied the correlation between the evolutionary curves of
the current-helicity density, computed from HMI vector magnetic
field, and the subsurface kinetic-helicity density, computed from
subsurface velocity field obtained from HMI time-distance
data-analysis pipeline, for two flare-productive active regions. For
the weighted helicities, we find that the evolution of the current
helicity and subsurface kinetic helicity has a high correlation,
larger than 0.60 for both active regions. But for the unweighted
case, essentially, both helicities do not show a high correlation
for the studied period for both regions. It is not clear why the
weighted parameters show a high correlation while the unweighted
parameters do not. Still, it is very interesting to see the weighted
helicities attain a high correlation during the evolution of both
active regions, especially when considering that these two
helicities are calculated from very different data. Although the
original observations were both from the same instrument HMI,
current helicity is calculated from vector magnetic field which was
derived from the 12-min-cadence Stokes parameters, and the
subsurface kinetic helicity is calculated from the 45-sec-cadence
Dopplergrams after some very complicated helioseismological
processing. The high correlation between these two helicities
indicate there are indeed some intrinsic connections between the
subsurface kinetic helicity and the photospheric current helicity.

Since the hemispheric preponderance of magnetic (or current)
helicity was reported \citep[e.g.,][]{Pevtsov95, BZ98}, many efforts
have been made to explain this observational phenomenon based on the
solar dynamo
\citep[e.g.,][]{Kleeorin03,Choudhuri04,Zhang06,Zhang12}, and the
$\Sigma$-effect of emerging magnetic flux \citep{lon98}. The adopted
velocity field here mainly comes from the region near the solar
surface, so the correlation between the short time scale variations
of kinetic and current helicity may be suitable to be explained by
using the $\Sigma$-effect. The $\Sigma$-effect suggested that during
the rising of the magnetic flux from convection zone to the
photosphere, the large-scale writhing and the local-scale twisting
together shape the tilt angle and magnetic (or current) helicity
distributions of active regions. The subsurface kinetic helicity
used in our analysis is presumably corresponding to the twisting
inside active regions in the upper convection zone. The in-phase
evolution of the subsurface kinetic helicity and the current
helicity probably indicate that the twisting beneath active region
surface indeed play an important role to shape the current helicity
distribution observed in the photosphere. On the other hand, perhaps
it is not surprising that these two helicities evolve in phase,
because beneath the photosphere magnetic field lines are frozen with
plasma and the evolution of photospheric magnetic field somehow
reflect the subsurface motions.

Another fact we cannot ignore is that despite the high positive
correlation in the evolutionary curves of the two helicities, the
signs of the two helicities do not often stay the same. In
particular, for AR~11158, the signs of the two helicities are more
often opposite than same. Therefore, although the evolution of the
two helicities is in positive correlation, it is quite likely that
we get a negative or no correlation if we pick just one random
snapshot. This may help explain why our previous study
\citep{GZZ09},  which statistically analyzed the correlation of the
two helicities using snapshots from many different active regions,
did not find a correlation. The results presented in this paper also
give us some guidance of how a statistical study with many active
regions should be carried out, and we plan to perform such a
statistical study once HMI accumulates sufficient number of active
regions.

\citet{Moll12} compared the flow structures in the upper
photospheric layers from their simulations and found that vertically
extended vortices were associated with magnetic flux concentrations.
Small-scale vortices occurring in simulations of non-magnetic or
weakly magnetized regions of the lower atmosphere was also studied
\citep{Kitiashivili12}. However, these simulations focused on
regions outside of sunspots with weaker magnetic field strength, and
the spatial scale was also smaller than the observational
resolution. Thus, we believe our study provides a useful window to
understand the temporal and spatial interactions between magnetic
field and flow field in a larger scale, which can be hardly achieved
using numerical simulations for the time being.

\begin{figure}
\epsscale{0.90} \plotone{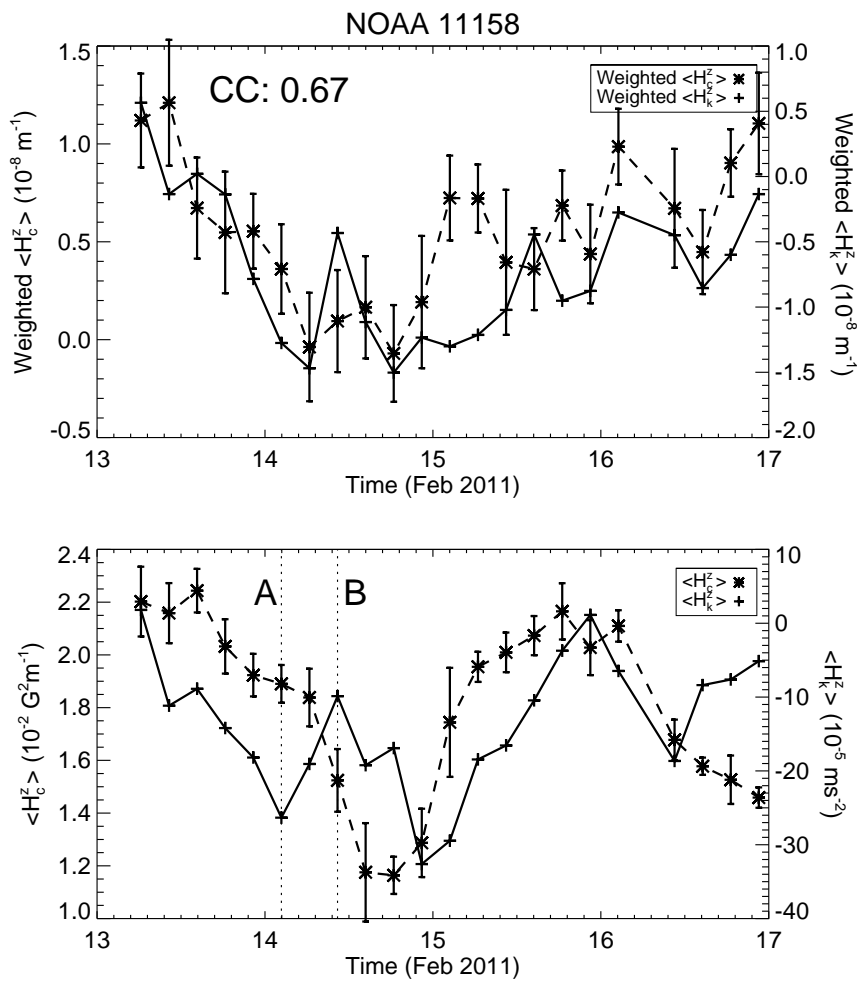}
\caption{{\it Top}: Evolution of the weighted current helicity
(marked with star, corresponding to the left vertical axis) and the
weighted subsurface kinetic helicity (marked with cross,
corresponding to the right vertical axis) for AR~11158. {\it
Bottom}: Same as the top panel but for the unweighted current
helicity and unweighted subsurface kinetic helicity. Dashed lines,
marked as ``A" and ``B", correspond to 10:00 UT and 14:00 UT of 2011
February 14, respectively.} \label{fig3}
\end{figure}

\begin{figure}
\epsscale{0.90} \plotone{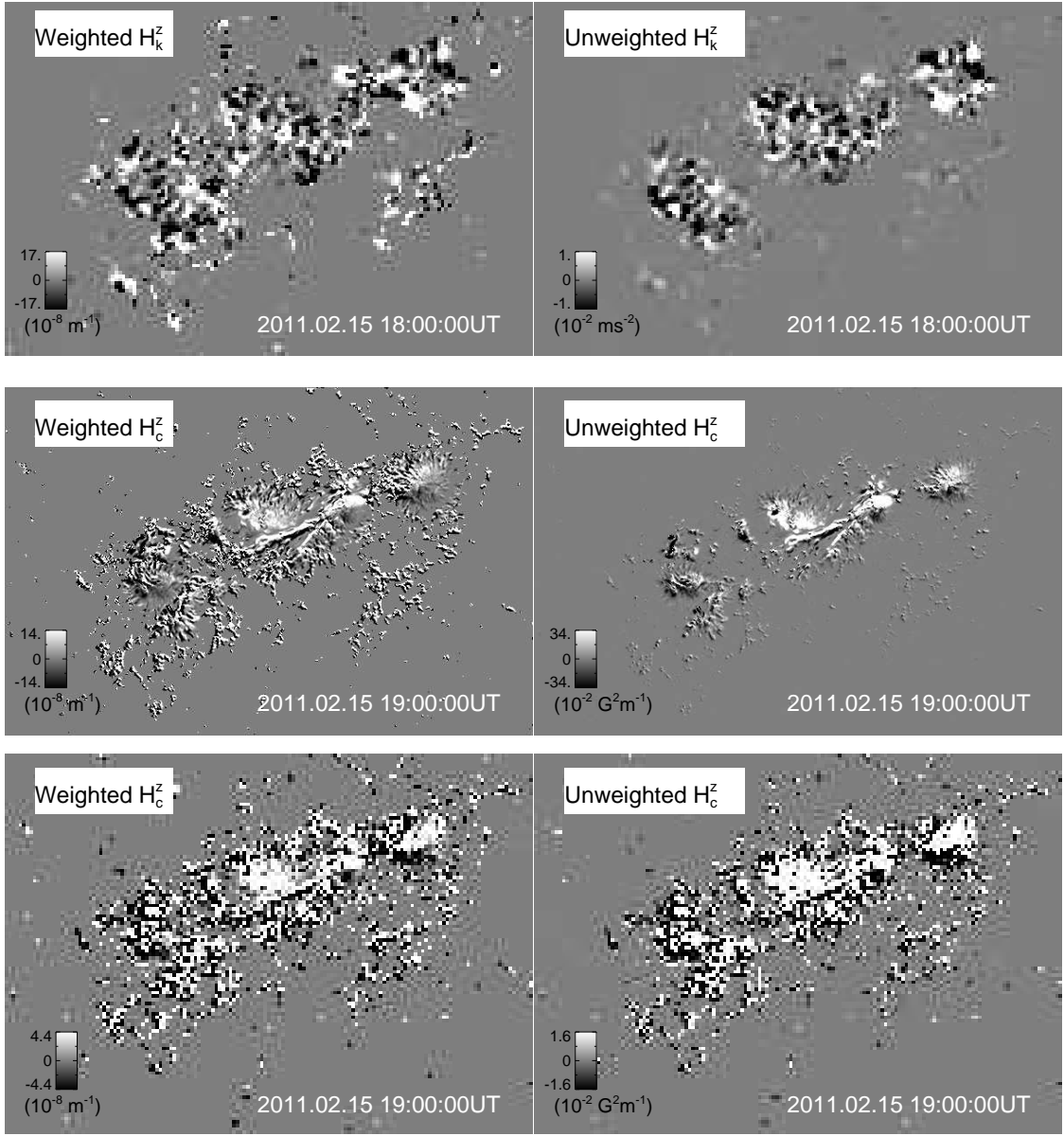} \caption{Snapshots showing
weighted $\langle H_k^z\rangle$ ({\it upper left}) and unweighted
$\langle H_k^z\rangle$ ({\it upper right}) at 18:00 UT of
2011.02.15, weighted $\langle H_c^z\rangle$ ({\it Middle left})
and unweighted $\langle H_c^z\rangle$ ({\it Middle right}) at 19:00
UT of 2011.02.15 with original spatial resolution, and weighted $\langle H_c^z\rangle$ ({\it Bottom left})
and unweighted $\langle H_c^z\rangle$ ({\it Bottom right}) after binning
down to match the spatial resolution of $\langle H_k^z\rangle$. The parameters are displayed relative to the corresponding standard deviations, and color bars at the left lower
corner of each panel show this scale. } \label{fig4}
\end{figure}

\begin{figure}
\epsscale{0.90} \plotone{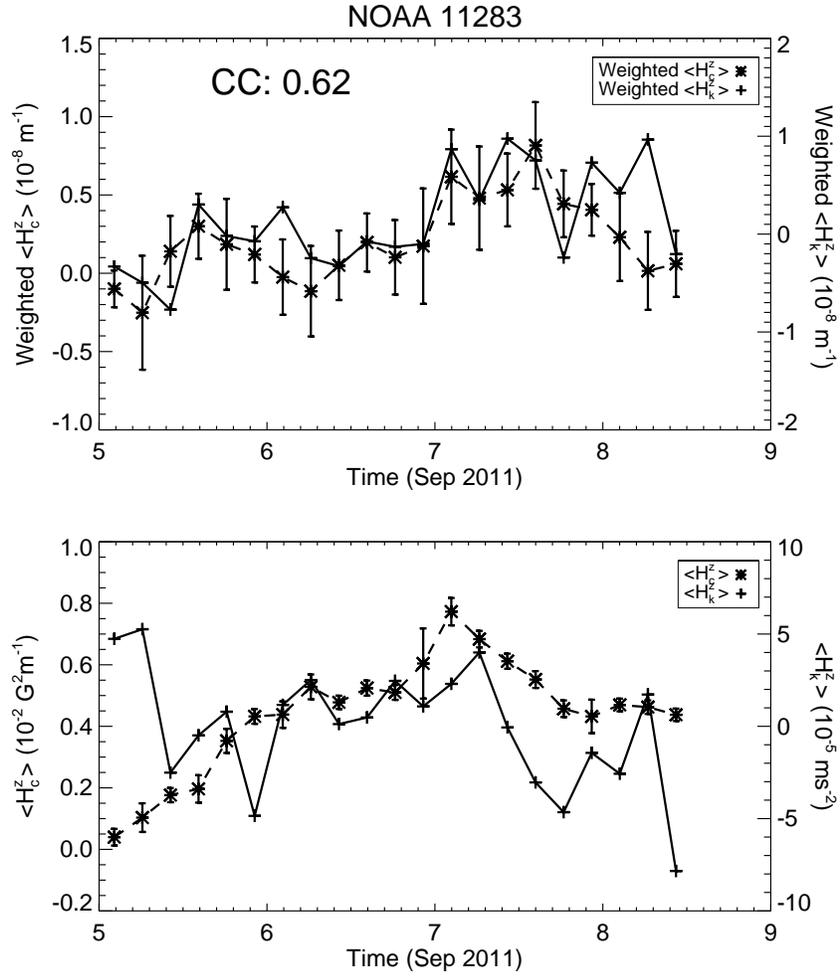}
\caption{Same as Figure~\ref{fig3}, but for NOAA AR~11283.}
\label{fig5}
\end{figure}

It is certainly of great importance to study whether there is any
indication of solar flare occurrences in the evolutionary curves of
both the current and subsurface kinetic helicities
\citep[e.g.,][]{Bao99,Komm05}. From the limited number of active
regions analyzed in this study, it is premature to draw a
conclusion. Figure~\ref{fig3}b seems to show that for AR~11158,
before the flare occurrence the subsurface kinetic helicity evolved
ahead of the current helicity, and evolved after the current
helicity after the flare. However, this phenomenon is not observed
in AR~11283. Nevertheless, this also needs a statistical study with
more active regions and more solar flares. The availability of
simultaneous HMI vector magnetic field and subsurface flow fields
for active regions makes such a study possible.

\acknowledgments SDO is a NASA mission, and HMI project is supported
by NASA contract NAS5-02139 to Stanford University. This work is
partially supported by the National Natural Science Foundation of
China under grants 11028307, 10921303, 11103037, 11173033, and
41174153,  and by Chinese Academy of Sciences under grant
KJCX2-EW-T07.

\clearpage

\clearpage

\clearpage

\clearpage

\clearpage

\end{document}